%% file: manuscript.tex
\documentclass[lettersize,journal]{IEEEtran}
\usepackage{amsmath,amsfonts}
\usepackage{algorithmic}
\usepackage{algorithm}
\usepackage{array}
\usepackage[caption=false,labelfont=rm,textfont=rm]{subfig}

\usepackage{textcomp}
\usepackage{stfloats}
\usepackage{url}
\usepackage{verbatim}
\usepackage{graphicx}
\usepackage{cite}
\begin{document}

\title{Decentralized Semantic Federated Learning for Real-Time Public Safety Tasks: Challenges, Methods, and Directions}

\author{Baosheng Li, Weifeng Gao, Zehui Xiong,~\IEEEmembership{Senior Member,~IEEE}, Jin Xie,~\IEEEmembership{Member,~IEEE}, \\ Binquan Guo and Miao Du
\thanks{Baosheng Li, Weifeng Gao (corresponding author), Jin Xie and Binquan Guo are with Xidian University, China; Zehui Xiong (corresponding author) is with Singapore University of Technology and Design, Singapore; Miao Du is with Southeast University, China}
}



\maketitle

\begin{abstract}
Public safety tasks rely on the collaborative functioning of multiple edge devices (MEDs) and base stations (BSs) in different regions, consuming significant communication energy and computational resources to execute critical operations like fire monitoring and rescue missions. Traditional federated edge computing (EC) methods require frequent central communication, consuming substantial energy and struggling with resource heterogeneity across devices, networks, and data.
To this end, this paper introduces a decentralized semantic federated learning (DSFL) framework tailored for large-scale wireless communication systems and heterogeneous MEDs. 
The framework incorporates a hierarchical semantic communication (SC) scheme to extend EC coverage and reduce communication overhead. Specifically, the lower layer optimizes intra-BS communication through task-specific encoding and selective transmission under constrained networks, while the upper layer ensures robust inter-BS communication via semantic aggregation and distributed consensus across different regions.
To further balance communication costs and semantic accuracy, an energy-efficient aggregation scheme is developed for both intra-BS and inter-BS communication. The effectiveness of the DSFL framework is demonstrated through a case study using the BoWFire dataset, showcasing its potential in real-time fire detection scenarios. Finally, we outlines open issues for edge intelligence and SC in public safety tasks.
\end{abstract}


\section{Introduction}
\IEEEPARstart{P}{ublic} safety tasks, such as fire monitoring, disaster response, and rescue missions, are critical operations that rely on the collaborative functioning of multiple edge devices (MEDs) and base stations (BSs) distributed across various regions \cite{1,2}. These tasks involve processing large volumes of real-time data collected from diverse sensors, drones, and other edge devices to support timely and effective decision-making in dynamic and often life-threatening environments. Seamless real-time data exchange and computational collaboration between MEDs and BSs are essential to ensure that these operations can adapt quickly to changing conditions, such as sudden environmental shifts \cite{2}.

However, the execution of real-time tasks is accompanied by significant challenges in terms of communication energy consumption and computational resource allocation. Traditional federated edge computing (EC) methods \cite{3}, including federated learning (FL) \cite{5,6}, are widely employed to address these needs but suffer from inherent limitations. Specifically, these EC methods \cite{3,5,6} rely heavily on frequent centralized communication with a central server for model exchange. This approach not only incurs substantial energy overhead but also struggles to cope with the heterogeneous nature of public safety networks, where devices, data distributions, and network conditions vary widely. For instance, MEDs in towns often experience unstable connectivity, limited bandwidth, and diverse computational capabilities.

\begin{figure}
    \centering
    \includegraphics[width=\linewidth]{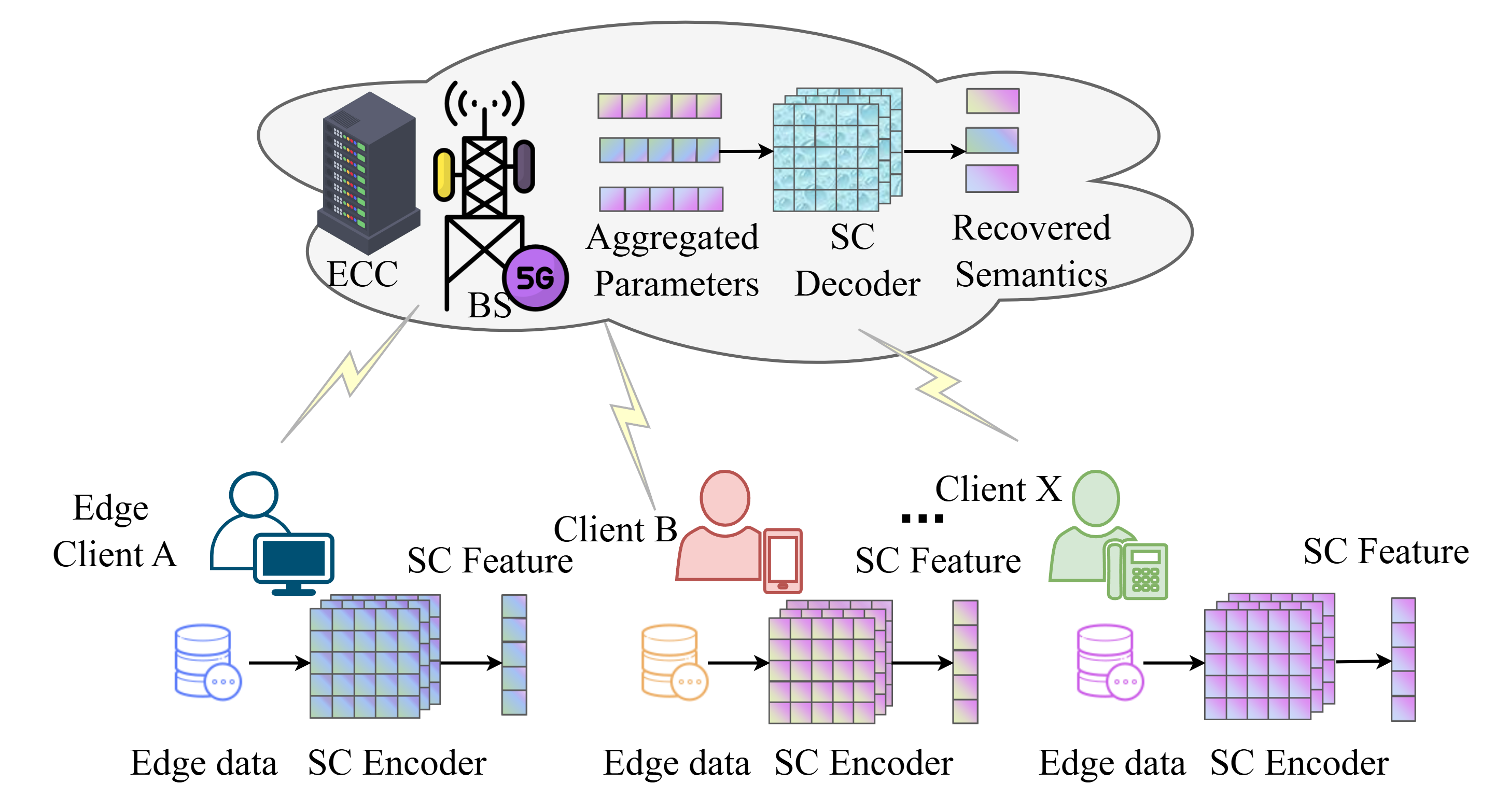}
    \caption{Integration of FL with SC in public safety tasks. The MEDs are used for semantic encoding, while the BS, functioning as a central server, aggregates the semantic data and facilitates data recovery. }
    \label{fig:1}
\end{figure}

To improve the efficiency and scalability of FL in public safety scenarios, current research leverages semantic communication (SC) technologies \cite{4,6,9}. By prioritizing meaningful data transmission, SC reduces communication overhead and enhances the ability of FL to support real-time, resource-constrained safety-critical applications. As illustrated in Fig. \ref{fig:1}, the BS, functioning as the edge computing center (ECC), is equipped with substantial computational and communication resources. By integrating SC into the FL framework, it is possible to concurrently perform efficient semantic transmission and carry out critical public safety tasks, enhancing both operational efficiency and task effectiveness \cite{4}. 
 Nevertheless, the dependence on a centralized BS introduces a single point of failure, rendering the system susceptible to disruptions \cite{5}. Additionally, communication bottlenecks and elevated latency can emerge when a large number of devices attempt to upload data concurrently, particularly in time-sensitive public safety tasks \cite{15}. Furthermore, the system's efficacy is highly contingent on network quality, which can deteriorate substantially in environments with limited connectivity, such as remote or underserved areas \cite{6,9}. 
To address these challenges, we propose an decentralized semantic federated learning (DSFL) for real-time public safety tasks. The contribution can be summarized as follows:

\begin{itemize}
    \item For large-scale wireless communication systems and heterogeneous MEDs environments, we develop a novel DSFL framework that effectively overcomes the high energy consumption and resource heterogeneity issues caused by frequent centralized communications in traditional EC methods.
    \item We propose a hierarchical SC framework. At the lower layer, task-specific encoding combined with selective transmission optimizes intra-BS communication efficiency. At the upper layer, semantic aggregation integrated with a distributed consensus mechanism ensures robust inter-BS data exchange, significantly reducing communication overhead and extending the operational range of edge computing networks.
    \item To optimize the trade-off between communication cost and semantic accuracy, we introduce an energy-efficient aggregation strategy that operates across both intra- and inter-BS contexts. The DSFL framework’s performance is validated through high-resolution fire image transmission and fire detection tasks, underscoring its practical viability and technical robustness in public safety applications.
\end{itemize}

\section{Challenges of FL in Public Safety} \label{sec1}
Real-time public safety tasks demand exceptional communication efficiency and computational power. In fire detection scenarios, for example, high-resolution images must be transmitted accurately and instantly so that edge-based artificial intelligence (AI) models can swiftly determine whether a fire is occurring. Although traditional EC methods like FL leverage the collective processing power of MEDs to train AI models collaboratively, they typically require abundant communication bandwidth and computing resources. With modern public safety evolving into large-scale, multi-regional monitoring under extreme communication conditions and involving heterogeneous MEDs, the resource-intensive nature of FL renders it increasingly impractical \cite{7}.

\subsection{Communication Challenges}
   \textbf{Network Unreliability:}
Public safety tasks often involve devices deployed in challenging environments, such as subway tunnels, remote regions, or disaster zones, where network conditions are poor or unstable. Network interruptions can delay or fail model synchronization, reducing the effectiveness of FL in critical scenarios.
Real-time public safety applications, like intelligent surveillance, require frequent communication between MEDs and servers to detect threats or respond to emergencies. FL mitigates communication bottlenecks by enabling local data processing and sharing only aggregated model updates. However, its performance still depends on stable network connectivity for synchronization. For example, delayed updates from MEDs in surveillance systems can hinder threat detection and emergency responses. It is crucial to ensure the effectiveness of FL in real-time public safety applications \cite{1}.

   \textbf{Resource Limitations:}
In public safety systems, MEDs must process and transmit high-definition images, audio, or video data, creating high demands on bandwidth and spectrum resources \cite{7, 9, 14}. FL relies on frequent communication for model updates, which adds further strain to these resources. Wireless technologies face challenges like channel congestion and interference when numerous devices compete for limited spectrum \cite{2}.
This issue becomes critical in emergency scenarios or densely populated areas, where real-time communication is essential. For example, in post-disaster situations, drones transmit high-definition images to aid rescue planning. While FL reduces the need to transmit raw data, limited bandwidth can still delay aggregated model updates, impacting timely decision-making. Such delays can hinder rescue efforts and reduce the overall effectiveness of public safety systems.
 These limitations highlight the importance of optimizing communication strategies in FL to ensure that it meets the stringent real-time requirements of public safety systems.

\subsection{Computation Challenges}
In public safety systems, FL must address the diversity of participating MEDs, which often have significant differences in data distributions, model architectures, communication protocols and computational power \cite{8}.

   \textbf{Heterogeneous Data:}
   MEDs participating in FL typically collect data from different scenarios or environments, leading to inconsistencies in the statistical properties of data distributions \cite{13}. This non-independent and identically distributed (Non-IID) nature makes it difficult for the global model to effectively learn from all MEDs. For example, in urban surveillance systems, cameras in different areas may face crowded zones, open spaces, or busy roads, resulting in video data with entirely different scene characteristics, exacerbating the Non-IID problem.

   \textbf{Heterogeneous Model Architectures:}
   Different MEDs may use entirely different model architectures, such as varying numbers of layers, types, and parameter scales in convolutional neural networks (CNNs). This heterogeneity in model architectures prevents devices from directly exchanging parameters or gradients. For instance, in disaster monitoring networks, some high-performance cameras may run complex CNNs (e.g., ResNet-152\footnote{https://huggingface.co/microsoft/resnet-152}), while low-performance sensors can only run lightweight models (e.g., EfficientNet-Lite\footnote{https://huggingface.co/docs/timm/models/tf-efficientnet-lite}). Directly sharing model parameters would lead to incompatibility issues \cite{10}.

   \textbf{Varying Communication Protocols:}
   MEDs may use different communication protocols (e.g., Wi-Fi, 4G, LoRa, satellite communication), with significant variations in speed, latency, and bandwidth capabilities, affecting the efficiency of FL \cite{13,14}. For example, in disaster rescue operations, drones may rely on satellite communication, while ground sensors use low-power network protocols. Incompatibility between these communication protocols can delay disaster prediction and emergency response.

   \textbf{Heterogeneous Devices:}
   The computational capabilities of MEDs participating in FL vary significantly, ranging from high-performance servers to low-power edge sensors \cite{13}. This hardware heterogeneity significantly impacts overall training efficiency and model quality. For instance, in intelligent security systems, surveillance devices in urban centers may connect to powerful edge computing nodes, while remote areas only have low-performance sensors, resulting in a significant gap in computational speed and training capabilities.

\section{DSFL Framework} \label{sec2}
\begin{figure*}[htbp]
    \centering
    \includegraphics[width=0.8\linewidth]{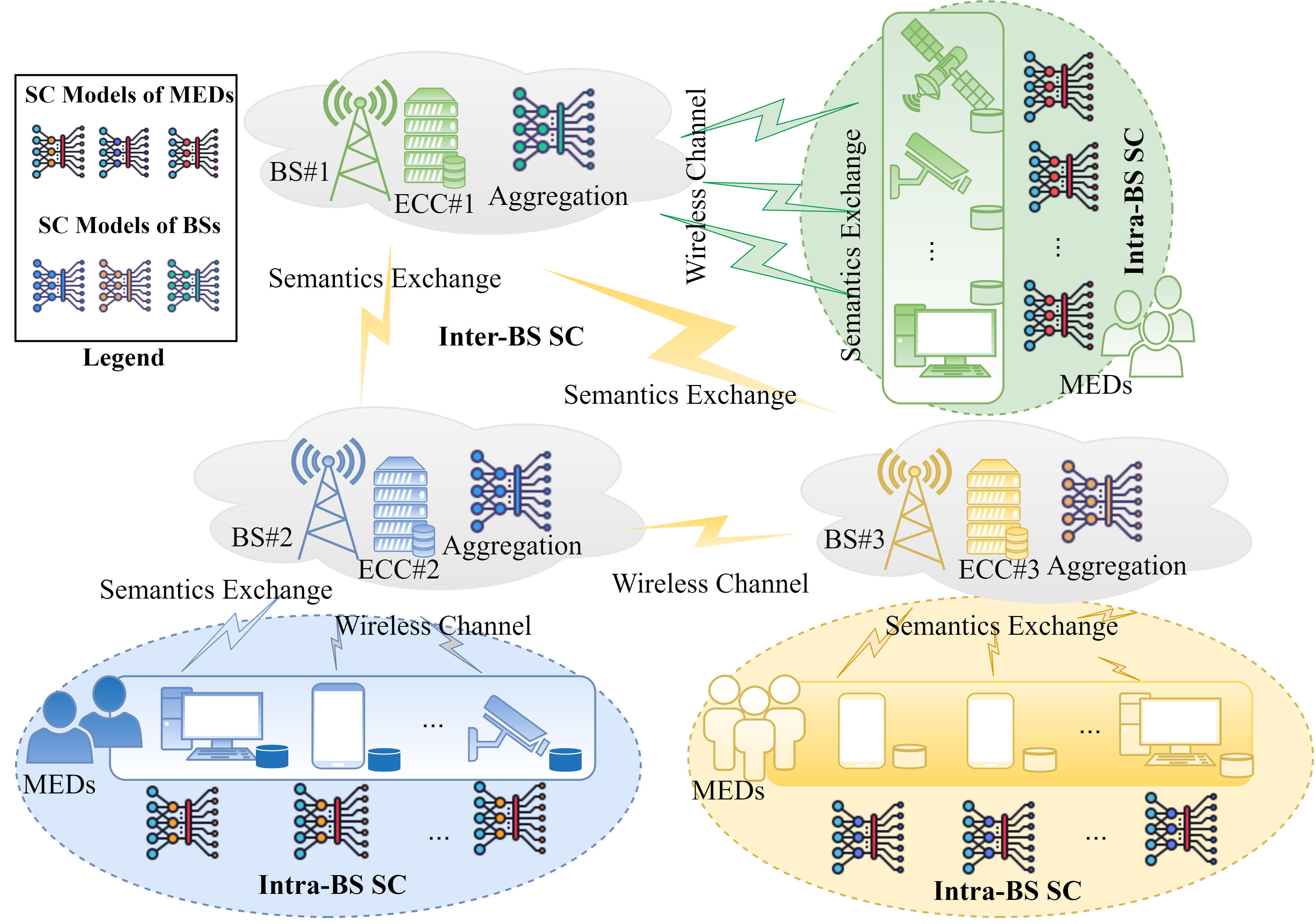}
    \caption{The framework of DSFL. It comprises MEDs and multiple BSs, organized into two layers: the lower layer enables intra-BS SC (BS-to-MED interactions within a coverage area), while the upper layer supports inter-BS SC (BS-to-BS coordination for global semantic fusion).}
    \label{fig:2}
\end{figure*}

In this section, we address the limitations of current FL frameworks when applied to public safety scenarios. As detailed in Section \ref{sec1}, FL frequently faces significant challenges, especially concerning communication efficiency and computational resource management, which are critical for handling real-time public safety tasks \cite{3,4,5,6}. This is particularly problematic in scenarios with a large number of devices, such as surveillance systems or emergency response networks, where transmission delays directly impact the timeliness and accuracy of safety-critical decisions. Hierarchical FL frameworks \cite{11} partially address these issues by introducing multi-level aggregation, where local updates are grouped and processed at intermediate nodes before being sent to a central server. Although Hierarchical FL reduces the frequency of global communication, it does not fully address the challenges of large-scale wireless communication and unstable networks. For instance, in public safety scenarios such as disaster zones or remote regions, network disruptions can hinder intermediate or global aggregation processes, leading to delays or failed updates. Additionally, these frameworks are often designed with static network conditions in mind, making them ill-suited for the highly dynamic and unpredictable environments.

To this end, we propose a novel DSFL framework, which incorporates a hierarchical SC scheme and a hybrid computational structure. Uniquely, it enables centralized federated semantic learning within the range of each BS and decentralized federated semantic transmission across multiple BSs. DSFL facilitates a seamless transition from local heterogeneous computing environments to a homogeneous computational framework on a global scale. Specifically, while the data stored on MEDs within a single BS may exhibit non-IID, the aggregation of data across different BSs transforms the distribution into IID. This is because the semantics aggregated at the BS level encompass a broader and more diverse range of information.

\subsection{Framework Overview}
Fig. \ref{fig:2} illustrates the comprehensive architecture of the DSFL framework, which is designed to address the challenges encountered in large-scale communication and EC within the realm of public safety. A detailed description of each component of the DSFL framework is provided as follows.

\textbf{Hierarchical SC Scheme:}
The hierarchical SC scheme combines centralized and decentralized synchronous communication for multi-BS and MEDs in public safety scenarios. Centralized synchronous communication within each BS ensures efficient aggregation of semantic information, while decentralized communication across BSs facilitates BS-to-BS exchange, promoting inter-region coordination and reducing reliance on a central server.
The hierarchical SC scheme can efficiently handle varying network reliability, device mobility, and fluctuating bandwidth availability, ensuring continuous and stable data exchange.

\textbf{Hybrid Computational Structure:}
The hybrid computational structure in DSFL integrates centralized FL within individual BSs and decentralized FL across multiple BSs. It allows local model training and semantic data processing at each BS, while facilitating the aggregation and exchange of semantic insights between BSs for global optimization. This structure balances local computation efficiency with cross-region coordination, making it well-suited for handling diverse public safety tasks.

\subsection{Local Semantic Model}
\begin{figure}[htbp]
    \centering
    \includegraphics[width=\linewidth]{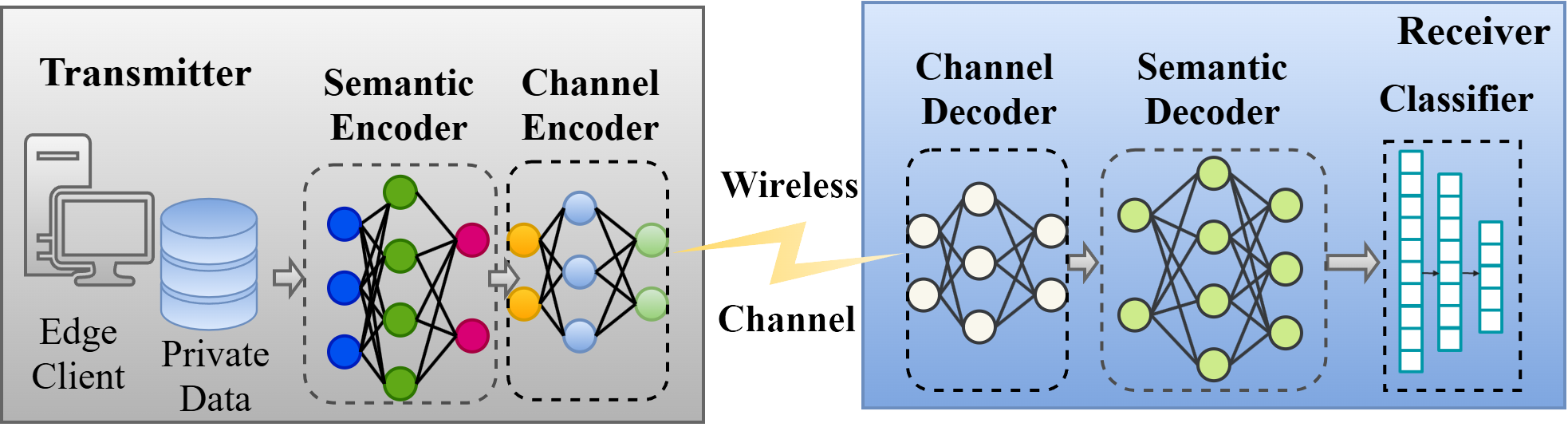}
    \caption{Local SC model in DSFL framework.}
    \label{fig:3}
\end{figure}

The local semantic model is designed to extract semantic features from raw data, aiming to reduce communication bandwidth usage and ensure the security of sensitive data. 
Fig. \ref{fig:3} illustrates the local semantic model, which consists of a transmitter and a receiver for extracting local semantics and receiving semantics from other devices. The SwinJSCC\footnote{https://github.com/semcomm/swinjscc} is used as the local semantic model of all MEDs.

\textbf{Transmitter:}
As a transmitter, each MED is responsible for gathering raw data from its surroundings, which may include but not limited to visual, auditory, and sensor-based information. To process this raw data and extract semantic meaning, the MED employs SwinJSCC encoder, a variant of vision transformer models. This model excels at identifying and extracting high-level features from complex datasets, such as images, thus transforming raw data into semantically rich representations that carry significant contextual information. Once the semantic information has been extracted, it needs to be prepared for transmission over the wireless channel.

\textbf{Wireless Channel:}
When transmitting images semantically over a wireless channel, the original high-level features extracted from the image data can suffer from several types of damage. Distortion may occur due to changes in signal amplitude or phase that do not faithfully represent the original information. Noise, on the other hand, introduces random variations into the signal, potentially obscuring the semantic content \cite{9}. To simulate the effects of noise encountered during real-world transmissions, additive white Gaussian noise (AWGN) is employed as a standard model. AWGN adds a constant level of white noise across all frequencies, which mimics the noise present in most wireless communication scenarios.

\textbf{Receiver:}
Upon reception of the wireless signal at the destination node, the first step is channel decoding, which demodulates the incoming signal to extract the embedded information. The output of this stage is a stream of digital symbols representing the encoded semantic information. Once the channel decoding has successfully demodulated the signal, the next critical phase is semantic decoding. At this point, the goal shifts from merely extracting the encoded data to restoring the high-level semantic features that were originally extracted by the transmitting device. We use SwinJSCC decoder to interpret the received digital symbols and reconstruct the intended semantic content. Finally, a classifier is employed to determine whether a public safety incident has occurred or the level of emergency.

\subsection{Energy-Efficient Aggregation}
\begin{figure*}
    \centering
    \subfloat[Intra-BS communication compression]{\includegraphics[width=\linewidth]{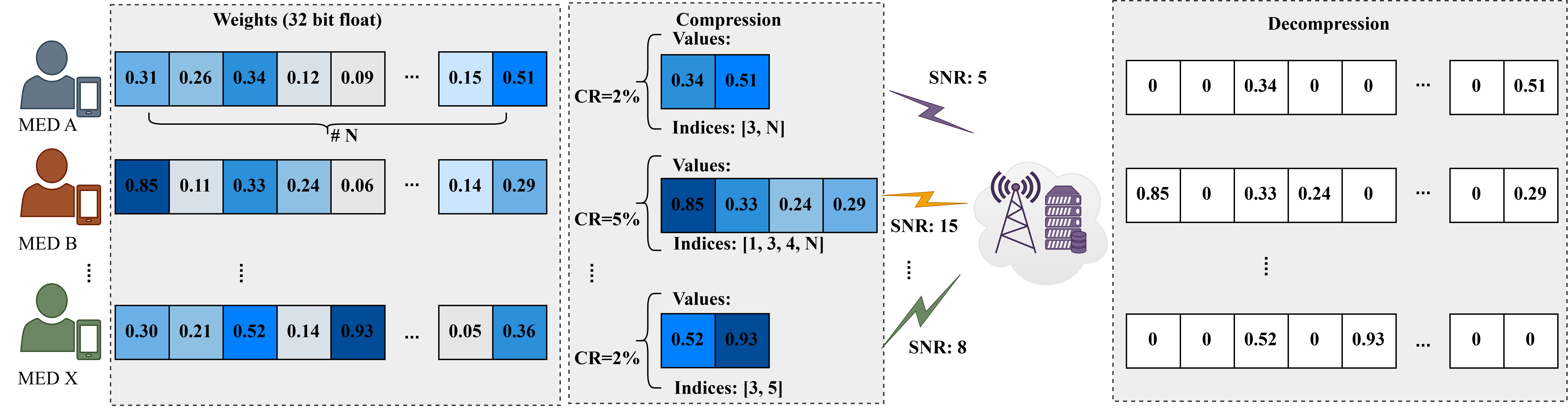}}\\
    \subfloat[Inter-BS communication compression]{\includegraphics[width=\linewidth]{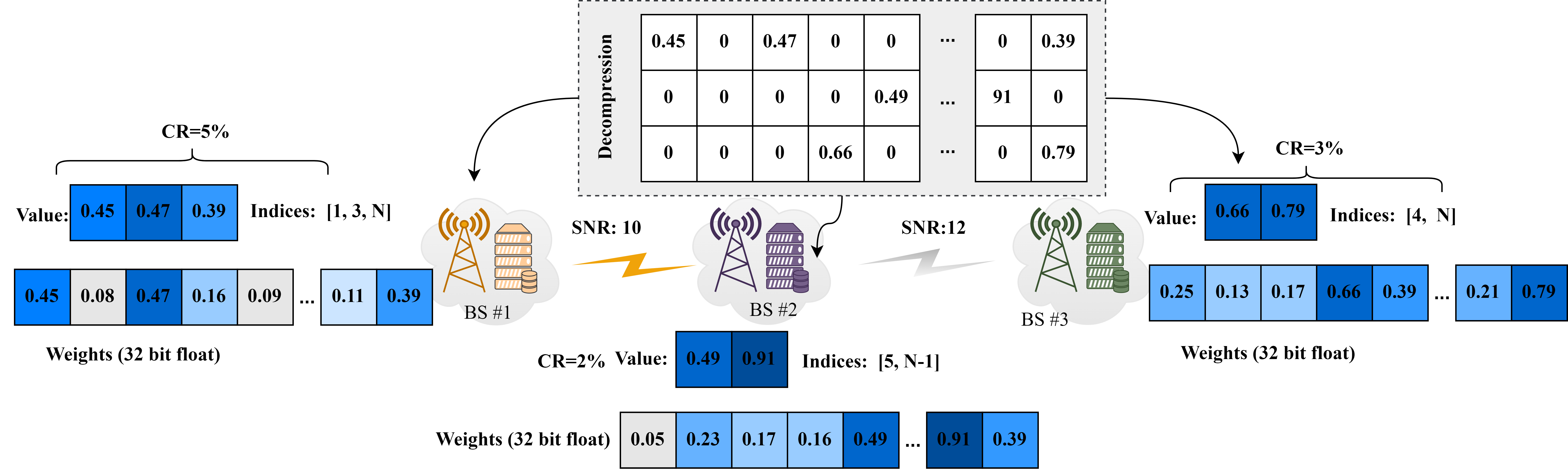}}    
    \caption{Energy-efficient compression schemes include intra-BS and inter-BS communication compression.}
    \label{fig:4}
\end{figure*}
Model aggregation is a pivotal process that ensures efficient and accurate information exchange between MEDs and BSs, as well as among BSs. 
The aggregation process is divided into two main phases: intra-BS aggregation and inter-BS aggregation. During the intra-BS aggregation phase, each BS collects semantic information from MEDs within its coverage area via wireless channels. The aggregated semantic content is then obtained by performing a weighted average of the received semantic information, where the weights can be determined based on factors such as signal quality or relevance of the data. In the inter-BS aggregation phase, each BS shares the aggregated semantic content with neighboring BS. This exchange allows for a broader consolidation of information across the network, facilitating more comprehensive insights and decision-making capabilities. However, the exchanged information includes the parameters of the semantic encoder, semantic decoder, channel encoder, and channel decoder, all represented as 32-bit floating-point values, resulting in a high transmission load. Additionally, frequent semantic exchanges within and between base stations can lead to substantial communication overhead, creating challenges in maintaining energy efficiency and ensuring operational reliability \cite{11}.
To address these challenges, we propose an energy-efficient aggregation scheme that optimizes both the precision of transmission and the energy expenditure. As shown in Fig. \ref{fig:4}, the energy-efficient aggregation scheme consists of two key steps:
\begin{itemize}
    \item[1)] \textbf{Intra-BS Compression}: Within each BS's coverage area, MEDs adaptively perform parameter compression based on the real-time Signal-to-Noise Ratio (SNR) between the MED and the BS. The compression rate (CR) dynamically adjusts according to the SNR. Specifically, CR decreases as the SNR increases, ensuring high transmission accuracy under favorable conditions; conversely, it increases as the SNR decreases, guaranteeing reliable transmission even under poor conditions while minimizing energy consumption. This adaptive strategy balances the need for precise data transfer with the constraints of limited power resources.
    \item[2)] \textbf{Inter-BS Compression}: During inter-BS aggregation, BSs apply the same compression strategy to their model parameters before transmitting them to neighboring BSs. This ensures that the aggregated models remain robust and up-to-date without imposing excessive loads on the network infrastructure.
\end{itemize}


\section{Case Study} \label{sec3}
In this section, we investigate the performance of the DSFL framework in the context of public safety tasks, specifically focusing on the SC accuracy for fire image transmission and the effectiveness of timely fire detection. The former is evaluated using the multi-scale structural similarity metric (MS-SSIM) and peak signal to noise ratio (PSNR), while the latter is assessed through detection accuracy and communication energy cost.
The BoWFire\footnote{https://bitbucket.org/gbdi/bowfire-dataset} dataset is employed, which comprises 226 images depicting industrial fires, traffic accidents, and harassment incidents. The maximum image size is 1056$\times$1024. These samples are randomly distributed across 20 MEDs, ensuring that each MED holds at least one data sample. 
Assume there are three BSs, with each BS area containing 1 to 10 MEDs.
Moreover, we employ the pre-trained SwinJSCC model as a foundation for fine-tuning in the task of fire image data transmission. Additionally, a fully connected neural network is utilized to perform fire detection based on the transmitted semantic information.
During the training process, we assume a network configuration comprising three BSs. The maximum transmission power for communication is set to 0.1 W. Both intra-BS and inter-BS SNRs vary dynamically between 0.1 dB and 20 dB, reflecting realistic wireless communication conditions. To manage the communication overhead and ensure efficient parameter exchange, we use the top-$k$ compression method for both MEDs and BSs. This technique allows only the most significant parameters to be uploaded, thereby reducing bandwidth requirements while maintaining accuracy. The number of local iteration is set to 5, meaning each MED will perform five rounds of local training before communicating. The global communication round is set to 100. For simulation purposes, we use an RTX 3090 GPU for the ECCs to accelerate computations. For the MEDs, we utilize an RTX 2080Ti to perform local training.

\begin{figure*}
    \centering
    \includegraphics[width=0.85\linewidth]{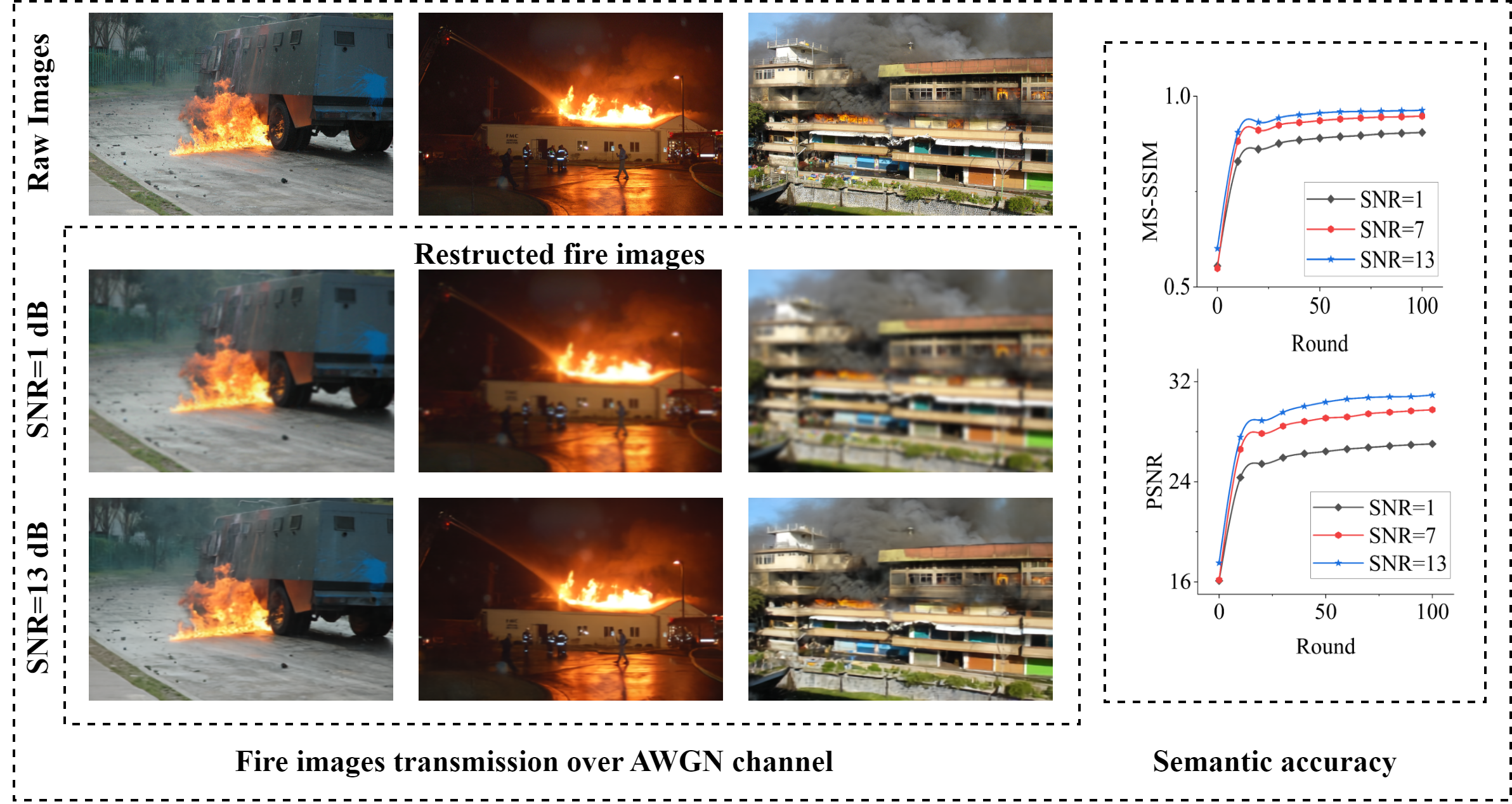}
    \caption{Performance comparison of the DSFL framework for transmitting fire images under varying SNR conditions. The left panel illustrates reconstructed images after transmission through channels with SNRs of 1 dB and 13 dB, while the right panel presents the corresponding quantitative metrics (MS-SSIM and PSNR) to evaluate transmission quality.}
    \label{fig:6}
\end{figure*}

The left side of Fig. \ref{fig:6} illustrates visual examples of fire images transmitted using the DSFL method under an AWGN channel with varying SNR conditions. At an SNR of 1 dB, the fire image reconstructed by the SwinJSCC decoder appears relatively blurry. However, as the SNR increases to 13 dB, the clarity of the decoded image improves significantly. This demonstrates the profound impact of SNR on the performance of DSFL.
\begin{figure}
    \centering
    \includegraphics[width=\linewidth]{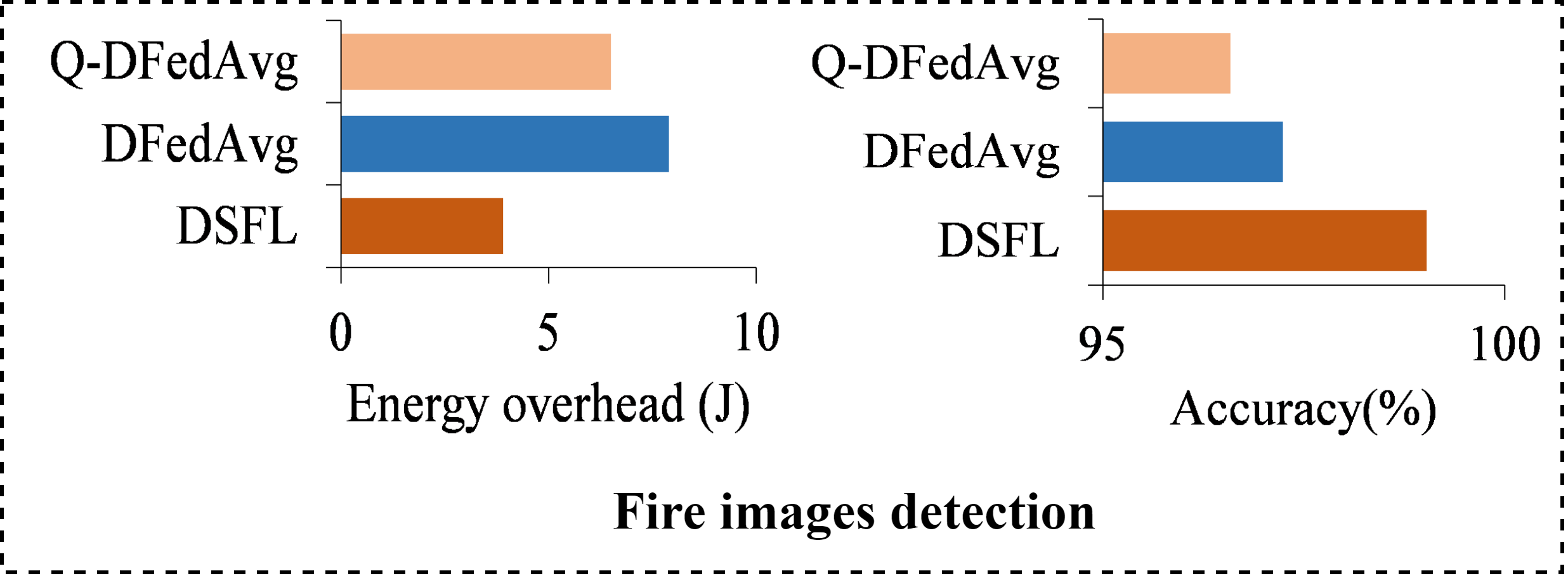}
    \caption{Performance comparison of the DSFL framework for fire images detection, along with an analysis of communication energy overhead. }
    \label{fig:7}
\end{figure}
The right side of Fig. \ref{fig:6} presents the convergence process in terms of MS-SSIM and PSNR under different SNR conditions. As the number of communication rounds increases, the MS-SSIM and PSNR between the decoded fire images and the original transmitted fire images progressively improve within the DSFL framework. Notably, the MS-SSIM and PSNR increase more rapidly when the SNR is 13 dB.
In addition, the energy consumption per communication round under the DSFL framework is of particular interest. Fig. \ref{fig:7} presents the comparison result with respect to the communication energy overhead and detection accuracy. Compared with DFedAvg and Q-DFedAvg \cite{13}, the results indicate that DSFL achieves lower energy consumption. This is attributed to the energy-efficient aggregation module in the DSFL framework, which adaptively compresses the transmitted parameters based on SNR, transmission power, and bandwidth. As a result, it effectively reduces the size of transmitted parameters from both MEDs and BSs, thereby lowering energy overhead.
Meanwhile, the detection accuracy of DSFL is the highest than other methods, which is due to the IID property from global scale.
Moreover, DSFL achieves superior accuracy in wildfire image detection by naturally converting Non-IID data into IID through a global perspective.

\section{Open Research Directions} \label{sec4}
Building upon the proposed DSFL framework, we outline three critical open research directions to further enhance edge AI and communication technologies for supporting rapid decision-making in public safety scenarios.

\subsection{Robust SC Scheme for DSFL in Public Safety}
One of the core challenges in deploying DSFL in public safety tasks is ensuring reliable and efficient communication in dynamic and often unstable network conditions. In emergency scenarios, such as during natural disasters or in remote areas, communication networks can be intermittent, making it difficult for MEDs and BSs to maintain consistent semantic information flow \cite{2,3}. A robust communication scheme can significantly enhance the DSFL framework by providing resilience to network disruptions, enabling low-latency communication, and optimizing bandwidth usage. This scheme would ensure that critical data, such as real-time emergency alerts or live video feeds, are reliably transmitted across MEDs and BSs, enabling rapid decision-making and coordinated response efforts.

\subsection{Multimodal SC for DSFL in Public Safety}
The data exchanged between MEDs is often highly heterogeneous. This includes various types of sensor data (e.g., video, audio, and text) that need to be effectively integrated for accurate decision-making. The DSFL framework already leverages SC for transmitting relevant information. However, multimodal semantic communication (MMSC) extends this capability by supporting multiple data modalities, enabling better contextual understanding across various sensor types \cite{8,10}. By incorporating MSSC, we can provide a richer, more comprehensive picture of the situation at hand. For example, combining video data from drones with environmental sensor readings can provide more accurate real-time insights for public safety operations.

\subsection{Privacy-Preserving Real-Time Edge AI}
Public safety applications often deal with sensitive information, such as personal data from surveillance cameras or GPS tracking devices \cite{0}. The DSFL framework inherently supports FL, which helps maintain data privacy by ensuring that raw data never leaves the local devices. However, further enhancing privacy-preserving capabilities is critical to ensure that real-time edge AI systems comply with stringent privacy regulations while still delivering high-quality insights \cite{5}. Research in this area can focus on developing advanced secure aggregation and differential privacy techniques to enhance the privacy guarantees while ensuring that MEDs can collaboratively learn and infer from data without compromising security and individual privacy.

\section{Conclusion}\label{sec5}
In this work, we proposed a novel DSFL framework for real-time public safety tasks. DSFL integrates a hierarchical SC scheme and a hybrid computational architecture, enabling it to address unstable communication and limited heterogeneous computing challenges in real-time public safety tasks. Additionally, the energy-efficient aggregation scheme allows DSFL to support operations under extreme communication scenarios.
Case study have demonstrated that DSFL achieves high semantic accuracy and energy-efficient communication while maintaining reliable task execution in challenging conditions. 
Future directions include exploring robust SC schemes, multimodal semantic processing, and privacy-preserving techniques.

 \input{reference.bbl}

 \bibliographystyle{IEEEtran}
\bibliography{reference}

 



\begin{IEEEbiographynophoto}{Baosheng Li}
(bs.li@stu.xidian.edu.cn) is currently pursuing a Ph.D. degree in Xidian University, China.
\end{IEEEbiographynophoto}
\vspace{-30pt}
\begin{IEEEbiographynophoto}
{Weifeng Gao}(gaoweifeng2004@126.com) received Ph.D. degree from Xidian University, China, in 2012. He is a full professor in Xidian University, China.
\end{IEEEbiographynophoto}
\vspace{-30pt}
\begin{IEEEbiographynophoto}
{Zehui Xiong}(zehui\_xiong@sutd.edu.sg) received the Ph.D. degree
in Nanyang Technological University (NTU), Singapore. He is an Assistant Professor
at Singapore University of Technology and Design,
and also an Honorary Adjunct Senior Research Scientist with Alibaba-NTU Singapore Joint Research
Institute, Singapore.
\end{IEEEbiographynophoto}
\vspace{-30pt}
\begin{IEEEbiographynophoto}
{Jin Xie}(xj6417@126.com) received the Ph.D. degree
in Xidian University, China. He is an Associate Professor
in Xidian University, China.
\end{IEEEbiographynophoto}
\vspace{-30pt}
\begin{IEEEbiographynophoto}
{Binquan Guo}(bqguo@stu.xidian.edu.cn) is currently pursuing a Ph.D. degree in Xidian University, China.
\end{IEEEbiographynophoto}
\vspace{-30pt}
\begin{IEEEbiographynophoto}
{Miao Du}(dumiao0118@seu.edu.cn)) is currently pursuing a Ph.D. degree in Southeast University, China.
\end{IEEEbiographynophoto}
\vfill

\end{document}